\documentclass[showpacs,preprintnumbers,twocolumn, pre]{revtex4-1}
\usepackage{graphicx}
\usepackage{amsmath}
\usepackage{amssymb}
\usepackage[latin1]{inputenc}
\newcommand{\be}{\begin{equation}}
\newcommand{\ee}{\end{equation}}
\newcommand{\ba}{\begin{eqnarray}}
\newcommand{\ea}{\end{eqnarray}}

\newcommand{\brt}{\begin{ruledtabular}}
\newcommand{\ert}{\end{ruledtabular}}

\usepackage{color,soul}
\setstcolor{red}

\begin{document}

\title{Evanescent continuous time random walks}

\author{E. Abad}
\email{Corresponding author: eabad@unex.es}
\affiliation{Departamento de F\'{\i}sica Aplicada, Centro Universitario de M\'{e}rida, Universidad de Extremadura,
E-06800 M\'erida, Badajoz, Spain}
\author{S. B. Yuste}
\affiliation{Departamento de F\'{\i}sica, Universidad de Extremadura, E-06071 Badajoz, Spain}
\author{Katja Lindenberg}
\affiliation{Department of Chemistry and Biochemistry and BioCircuits Institute, University of
California San Diego, La Jolla, California 92093-0340, USA}

\begin{abstract}
We study how an evanescence process affects the number of distinct sites visited by a continuous time random walker in one dimension.  We distinguish two very different cases, namely, when evanescence can only occur concurrently with a jump, and when evanescence can occur at any time. The first is characteristic of trapping processes on a lattice, whereas the second is associated with spontaneous death processes such as radioactive decay. In both of these situations we consider three different forms of the waiting time distribution between jumps, namely, exponential, long-tailed, and ultra-slow.
\end{abstract}

\pacs{05.40.-a, 82.20.Fd}

\maketitle

\section{Introduction}

One of the most thoroughly studied properties of random walkers is the territory explored as the walk proceeds.  In the traditional walk taking place on a lattice on which the walker moves by taking steps from one lattice site to another, this territory is measured by counting the number of sites  visited after $n$ steps.  In particular, it is the average of this quantity over an ensemble of walkers, which we will denote by $ S_n$, that has been most extensively studied (see e.g. \cite[chapter 6]{Hughes}). The asymptotic ($n\to\infty$) results $S_n  \propto n^{1/2}$, $\propto n/\ln(n)$, and $\propto n$ in one, two, and three dimensions respectively are then used to calculate other important related quantities such as the asymptotic survival probability of a target surrounded by diffusing traps.  Note that the $S_n$ all diverge as $n\to\infty$. The transcription to continuous time is accomplished by going from $n$ to $t$ using the waiting time distribution. In continuous space and time the results for the average number of distinct sites visited can be translated to the average volume of the Wiener sausage generated up to time $t$ with an appropriate transcription of sites and steps to volume and time \cite{Spitzer1964, Berezhkovskii1989}.

Our interest lies in the calculation of these quantities, but now when the walker can instantaneously die in the course of the motion, at which point no new sites (or volume) are visited (we use ``evanesce'' as a synonym of ``die'' throughout the paper).  In recent work \cite{ourPRL, MortalBookChapter} we carried out these calculations for a random walker (or a diffusing particle) that can evanesce either exponentially, typical of a unimolecular decay, or according to a power law, indicative of a more complex evanescence process.  We discussed the consequences of the modifications introduced by the evanescence process in exploration properties of random walkers and related quantities such as the survival probability of a target surrounded by evanescent diffusing particles \cite{SurvivalEvanescent, AYLJMMNP}.  In this work we extend those results to random walkers whose stepping statistics may not be those of a Markovian random walk or, equivalently, of a simple diffusion process. Thus, we now deal with a more general continuous time random walk (CTRW) of evanescent walkers.  We consider three examples of waiting time distributions for our walkers: exponential, long-tailed, and ultra-slow. The first one is associated with a linear time dependence of the mean square displacement at long times, associated with ordinary diffusion, whereas the latter two result in subdiffusive behavior.

We also differentiate walkers according to a second characteristic, namely, \emph{when} exactly the walker may evanesce.   In one case we assume that death can only occur at the moment of stepping. In contrast, in the second case we decouple the two processes, the walking and the evanescing, so that the walker can die at any moment. The first situation can be regarded as characteristic of trapping processes on the lattice. An example is the case of walkers that have a finite probability of being absorbed or converted into an inert species by the substrate every time they step from one site to another (see e.g. Sec. 5 in \cite{WalshKozak}).  Another example occurs on surfaces where some jumps lead to irreversible escape from the surface. These can formally be treated as a reaction resulting in particle disappearance at the time of a jump \cite{Shkilev2011}. On the other hand, the decoupled case corresponds to particle disappearance processes which do not depend on transport properties, such as for instance radioactive decay of a diffusing isotope \cite{Zoia2008}. In this context, the extent of the region contaminated by the isotope before decaying into a stable species (tantamount to the number of distinct sites visited) is of special interest.

Interestingly, there are cases where the precise nature of the coupling between evanescence processes and transport is unknown. In some of these it has been established that the above two choices may lead to entirely different results \cite{HLWPRE06}, leading to the possibility of experimental determination of the coupling.  One such example occurs in the formation of morphogen gradients (morphogens are special signaling molecules of importance in embryogenesis): In ref. \cite{HornungPRE05}, irreversible degradation (``death'') of subdiffusive morphogens was assumed to occur only when the morphogens perform a transition between one binding site and the next, leading to the absence of stationary morphogen concentration gradients in the long time limit. By way of contrast, in ref. \cite{YusteAbadLindenberg10} we assumed that degradation could occur at any time during transport, resulting in the formation of a stationary gradient.

We organize our paper as follows.  In Sec.~\ref{Markovian} we  introduce  exponential evanescence in the context of an ordinary random walk (Markovian case), so that our new results can be seen in the proper context. In Sec.~\ref{CTRW} we extend the analysis to CTRWs with different waiting time distributions: exponential, long-tailed (inverse power law decay with time), and ultra-slow (inverse power law decay with the logarithm of time).  In Sec.~\ref{associated} we consider these three cases when evanescence events can only occur concurrently with a step, and in Sec.~\ref{notassociated} when evanescence can occur at any time independently of the stepping process.  In Sec.~\ref{conclusions} we summarize our results and conclude with some additional observations.

\section{Markovian case with exponential evanescence}
\label{Markovian}

We recently dealt with the problem of the distinct number of sites visited by an ensemble of ordinary random walkers that evanesce according to the exponential decay \cite{ourPRL, MortalBookChapter}
\be
\rho(n)=\exp(-\lambda n)
\label{rhon}
\ee
for the surviving fraction of walkers (or the probability that a walker survives after $n$ steps). We do not repeat the calculation done earlier except to provide some results relevant for this work.
Specifically, we consider an ordinary nearest neighbor random walk.
Let $P_n(s|s_0)$ be the probability of finding a non-evanescent walker at site $s$ after $n$ steps given that it started at site $s_0$ at step $n=0$, and let $\widehat P(s|s_0;\xi)$ be the generating function for this probability,
\be
\widehat P(s|s_0;\xi) = \sum_{n=0}^\infty P_n(s|s_0) \xi^n.
\label{genfcn}
\ee
Let $S_n$ denote the average number of distinct sites visited by an ensemble of non-evanescent random walkers and $S_n^*$ the same average but for evanescent walkers (the asterisk is used to indicate that the relevant quantity refers to evanescent walkers).  The generating functions for $S_n$ and  $S_n^*$ are defined as
\begin{equation}
\widehat S(\xi) = \sum_{n=0}^\infty S_n \xi^n, \qquad \widehat S^*(\xi)= \sum_{n=0}^\infty S_n^* \xi^n .
\label{generating.functions}
\end{equation}
We showed \cite{ourPRL} that these generating functions are related to the generating function $\widehat P(s_0|s_0;\xi)$ for the probability of return to the origin of a non-evanescent random walker as follows:
\begin{eqnarray}
\widehat S(\xi)&=& \frac{1}{(1-\xi)^2} \frac{1}{\widehat P(s_0|s_0;\xi)},
\label{alive}
\\
\widehat S^*(\xi)&=& \frac{1}{(1-\xi)}\; \frac{1}{(1-e^{-\lambda} \xi)} \frac{1}{\widehat P(s_0|s_0;e^{-\lambda}\xi)}.
\label{S*xi}
\end{eqnarray}
The discrete Tauberian theorem (see e.g. \cite[p. 118]{Hughes}) for the non-evanescent case yields the asymptotic results mentioned in the previous section for the average number of distinct sites visited,  all of which diverge as $n\to\infty$, while for the evanescent problem we find the \emph{finite} value
\be
S^*_\infty=\frac{1}{(1-e^{-\lambda})} \frac{1}{\widehat P(s_0|s_0;e^{-\lambda})}.
\ee
In dimension $d=1$ we have the well-known result
\be
\label{1dgenfunc}
\widehat P(s_0|s_0;\xi)=\sum_{n=0}^\infty  \; P_n(s_0|s_0)\; \xi^n=(1-\xi^2)^{-1/2},
\ee
whence
\be
\widehat S^*(\xi) = \frac{(1+e^{-\lambda}\xi)^{1/2}}{(1-\xi)(1-e^{-\lambda}\xi)^{1/2}}
\label{explicit}
\ee
and
\be
S^*_\infty=\left(\frac{1+e^{-\lambda}}{1-e^{-\lambda}}\right)^{1/2}.
\label{S*inf}
\ee

In addition to the asymptotic finite average number of distinct sites visited, it is possible to calculate the subdominant term that describes the long time ($n\to \infty$) approach to the asymptotic result. We find \cite{MortalBookChapter}
\be
\label{SnLargen}
  S_n^*   \sim S_\infty^*   - \sqrt{2}\,
\frac{ I_{e^{-\lambda}}(n+1,1/2)}{(1-e^{-\lambda})^{1/2}},
\ee
where $I_x(a,b)$ stands for the regularized Beta function.
This result agrees with numerical simulation results even for rather moderate values of $n$ \cite{MortalBookChapter}. For small values of $\lambda$ one gets the simpler expression
\be
S_n^* \sim  \sqrt{\frac{2}{\lambda}}
- \frac{1}{\lambda}\, \sqrt{\frac{2}{\pi n}} \,
 e^{-\lambda n}  .
\label{traditional}
\ee

\section{CTRW walk in one dimension with exponential evanescence}
\label{CTRW}

A nearest neighbor CTRW is defined by the distribution of stepping times $\psi(t)$.  This is the probability density that a step occurs exactly $t$ time units after the previous step.  If the time $t=0$ is chosen to coincide with a step, then we can implement the recursion relation \cite[p. 96]{BookWeiss}
\be
\label{convpsi}
\psi_{n+1}(t)=\int_0^t \psi_n(t')\,\psi(t-t')\,dt'
\ee
for the probability density $\psi_n(t)$ that step $n$ occurs exactly at time $t$.
If $t=0$ is not a stepping time then one has to treat the first step differently.  We will henceforth assume that $t=0$ is a stepping time to avoid a further complication that can be addressed using well-known methods \cite{FirstStepsBook},
but that does not add much to the points we wish to highlight
 in this paper. The Laplace transform of this convolution clearly leads to \cite{BluKlaZuOptical, BlumenEtAl84}
\be
\widetilde{\psi}_n(u)=\left [ \widetilde{\psi}(u)\right ] ^n.
\ee
Here a tilde indicates a Laplace transform with respect to time and $u$ is the transform variable.

The time $t$ may not be exactly a stepping time, that is, the $n$-th step might occur at a time $t'< t$.  Thus, the probability that exactly $n$ steps have been executed by the walker by time $t$ is
\be
\label{intchi}
\chi_n(t)=\int_0^t \psi_n(t')\,\Phi(t-t')\,dt',
\ee
where $\Phi(t-t')=1-\int_{0}^{t-t'} \psi(t'')\,dt''$ is the probability that the walker has not moved in the time interval $t-t'$ since the $n$-th stepping time $t'$. It then follows that the Laplace transform of $\chi_n(t)$ is
\be
\widetilde{\chi}_n(u) = \frac{1-\widetilde{\psi}(u)}{u}\left [ \widetilde{\psi}(u)\right] ^n,
\label{familiar}
\ee
a familiar random walk result \cite{DistinctSitesCTRW}. We go on to use this result in the next subsection.

\subsection{Evanescence associated with steps}
\label{associated}

Consider the case where evanescence can only occur concurrently with a jump and not when the walker is standing still waiting between jumps.
The average number of distinct sites visited by an ensemble of evanescent walkers up to time $t$, $S^*(t)$, is then given by the decomposition
\be
\label{s*t}
S^*(t) =\sum_{n=0}^\infty  S^*_n \chi_n(t),
\ee
with a similar decomposition for $S(t)$ in terms of $S_n$ for non-evanescent walkers.  $\chi_n(t)$ is the probability that a \emph{non-evanescent} walker has jumped exactly $n$ times between times $0$ and $t$. The information about the evanescence is thus entirely contained in $S_n^*$.   It follows that the Laplace transform of $S^*(t)$ then is
\be
\label{lpS*}
\widetilde{S}^*(u)=\frac{1-\widetilde \psi(u)}{u}\sum_{n=0}^\infty [\widetilde{\psi}(u)]^n \, S^*_n = \frac{1-\widetilde{\psi}(u)}{u} \,  \widehat S^*\left( \widetilde{\psi}(u)\right),
\ee
completely analogous to the familiar expression for non-evanescent walkers \cite{DistinctSitesCTRW}.  Using the explicit form \eqref{explicit} of the generating function we obtain
\be
\label{s*u}
\widetilde{S}^*(u)=\frac{1}{u}\, \frac{[1+e^{-\lambda}\widetilde{\psi}(u) ]^{1/2}}{[1-e^{-\lambda}\widetilde{\psi}(u)]^{1/2}}.
\ee

Note that the time dependence of $\rho(t)$, the probability that a walker survives up to time $t$, depends on how many steps a walker has taken up to that time, and is thus dependent on the waiting time distribution $\psi(t)$.
We now go on to implement these results for a variety of waiting time distributions.

\subsubsection{Exponential waiting time distribution}

First we consider the waiting time distribution associated with ordinary diffusion, namely, an exponential,
\be
\psi(t)=\omega \,e^{-\omega t}.
\label{exponentialwtd}
\ee
The mean waiting time is $T=\int_0^\infty t\,\psi(t)\,dt=\omega^{-1}$, that is, $\omega$ is the mean number of steps per unit time.  The Laplace transform of the waiting time distribution then is $\widetilde{\psi}(u)=\omega/(\omega+u)$, so that
\be
\widetilde{S}^*(u)=\frac{1}{u}\,\frac{[u+(1+e^{-\lambda})\omega]^{1/2}}{[u+(1-e^{-\lambda})\omega]^{1/2}}.
\ee
This result can be analytically inverted \cite[p. 210, formula 40]{RobertsKaufman}
 to obtain
\begin{eqnarray}
S^*(t)&=&e^{-\omega\,t}\,I_0(e^{-\lambda}\omega\,t)\nonumber\\
&&+\omega (1+e^{-\lambda})\int_0^t e^{-\omega \, t'}\,
I_0(e^{-\lambda}\omega\,t')\,dt',
\label{s*t2}
\end{eqnarray}
where $I_0(\cdot)$ is a modified Bessel function.  When $t\to\infty$, the first term on the right vanishes, and we are left with \cite[p. 708,  formula 4]{GradshteynRyzhik})
\begin{eqnarray}
S^*(\infty)&=&\omega (1+e^{-\lambda})\int_0^\infty e^{-\omega \, t'}\,
I_0(e^{-\lambda}\omega\,t')\,dt'\nonumber\\
&=&\left( \frac{1+e^{-\lambda}}{1-e^{-\lambda}}\right)^{1/2}.
\label{asymp1}
\end{eqnarray}
This result agrees with Eq.~(\ref{S*inf}), as it should, since a CTRW with exponential stepping times is equivalent to an ordinary random walk in the long time limit.

It is useful for later comparison to calculate the way in which the asymptotic result for the distinct number of sites visited is approached at long times.  This can be done by rewriting Eq.~(\ref{s*t2}) as follows:
\begin{eqnarray}
S^*(t)&=&S^*(\infty)- \omega (1+e^{-\lambda})\int_t^\infty e^{-\omega \, t'}\;
I_0(e^{-\lambda}\omega\,t')\,dt' \nonumber\\
&&+e^{-\omega\,t}\,I_0(e^{-\lambda}\omega\,t).
\end{eqnarray}
Next we implement the asymptotic expansion for large $|z|$,
\be
I_\nu(z)\sim \frac{e^z}{\sqrt{2\pi z}}\left\{1-\frac{4\nu^2-1}{8z}
+\ldots\right\}.
\ee
Keeping only the leading long-time contribution to the asymptotic expansion, we see that
\begin{eqnarray}
&&\int_t^\infty e^{-\omega \, t'}\,I_0(e^{-\lambda}\omega\,t')\,dt'\nonumber\\
&& \quad\sim \omega^{-1}\,\frac{1}{1-e^{-\lambda}}
\frac{\exp{[\displaystyle( e^{-\lambda}-1)\,\omega t]}}{\sqrt{2\pi e^{-\lambda}\omega t}} ,
\end{eqnarray}
and keeping only the leading long-time contribution to the asymptotic expansion, we find
\begin{eqnarray}
\label{asymptS}
S^*(t)&\sim&S^*(\infty) \nonumber\\
&&- \frac{2e^{-\lambda}}{1-e^{-\lambda}} \frac{\exp{[\displaystyle(
e^{-\lambda}-1)\,t/T]}}{\sqrt{2\pi e^{-\lambda} t/T}}\, .\nonumber\\
\end{eqnarray}
We see that the distinct number of sites visited by our CTRWer with an exponential stepping time distribution and exponential evanescence, with evanescence occurring only at the same time as a jump and not while the walker is waiting, is finite, and that the approach to the asymptotic result is faster than exponential.
When $\lambda \to 0$  the above result simplifies to
\be
\label{slreacS}
S^*(t)\sim\sqrt{\frac{2}{\lambda}}-\frac{1}{\lambda}\sqrt{\frac{2}{\pi \omega t}}\,e^{-\lambda \omega t},\qquad t\to\infty.
\ee
Equation~(\ref{traditional}) with $S_\infty^*$ given explicitly in the small $\lambda$ limit is in turn given by
\be
S_n^* \sim\sqrt{\frac{2}{\lambda}}-\frac{1}{\lambda}\sqrt{\frac{2}{\pi n}}\,e^{-\lambda n}, \qquad n\to\infty.
\ee
The two results thus coincide if one sets $n=t/T=\omega t$.

Finally, since the results are handy, we quickly compare the CTRW and the Markovian random walk models when there is no evanescence (and the results given in the introduction hold). The distinct number of sites visited as $n\to\infty$ or as $t\to\infty$ of course diverge. Using Eq.~(\ref{1dgenfunc}) in Eq.~(\ref{alive}) and implementing the exponential waiting time distribution, we have
\be
\widetilde{S}(u)=\frac{1}{u}\,\sqrt{\frac{u+2\omega}{u}},
\ee
whose inverse Laplace transform is known exactly \cite[p. 210, formula 34]{RobertsKaufman}:
\be
S(t)=e^{-\omega\,t}\left[(1+2\omega t)\,I_0(\omega t)+2\omega t\,I_1(\omega t)\right].
\ee
It is not difficult to show that this result coincides with Eq.~(\ref{s*t2}) when the limit $\lambda \to 0$ is taken in the latter. Asymptotically we find
\be
S(t)\sim \sqrt{\frac{8\omega\,t}{\pi}}, \quad t\to\infty,
\label{classic1d}
\ee
which is the classic result for a random walk in one dimension if again we set $\omega t=t/T=n$. Note that the limit $\lambda \to 0$ of Eq.~(\ref{slreacS}) does not give Eq.~(\ref{classic1d}), i.e., the case $\lambda = 0$ is singular.

\subsubsection{Long-tailed waiting time distribution}

Next, while still considering exponential evanescence and evanescence events that only occur concurrently with a jump, we turn to CTRWs with a long-tailed waiting time distribution. Specifically, at long times we consider a waiting time distribution that decays as a power law \cite{DistinctSitesCTRW},
\be
\psi(t)\sim \frac{\gamma \tau^\gamma t^{-1-\gamma}}{\Gamma(1-\gamma)}, \qquad t\to\infty,
\label{longtail}
\ee
where $0<\gamma<1$ and $\tau>0$.  The Laplace transform of $\psi(t)$ for small $u$ is
\be
\label{asymptpsiu}
\widetilde{\psi}(u) \sim 1-\tau^\gamma u^\gamma, \qquad u\to 0.
\ee
Using this expression in Eq.~\eqref{s*u} in turn gives
\be
\label{Sast24}
\widetilde{S}^*(u)\sim \frac{1}{u}\,\frac{[1+e^{-\lambda}-e^{-\lambda}\tau^\gamma u^\gamma]^{1/2}}
{[1-e^{-\lambda}+e^{-\lambda} \tau^\gamma u^\gamma]^{1/2}}, \qquad u\to 0.
\ee
Taking into account that
\be
\label{exp1}
\frac{(\alpha- z)^{1/2}}
{(\beta+ z)^{1/2}}\sim \left(\frac{\alpha}{\beta}\right)^{1/2}\hspace{-3mm} -\,\frac{ \alpha+\beta }{2(\alpha \beta^3)^{1/2}}\,z, \qquad z\to 0,
\ee
Eq.~\eqref{Sast24} yields
\be
\widetilde{S}^*(u)\sim \left( \frac{1+e^{-\lambda}}{1-e^{-\lambda}}\right)^{1/2}\hspace{-3mm}u^{-1}
-\frac{e^{-\lambda}\tau^\gamma}{(1+e^{-\lambda})^{1/2}(1-e^{-\lambda})^{3/2}}\,u^{\gamma-1}
\ee
as $u \to 0$. Laplace inverting each term finally gives us the result
\be
S^*(t)\sim
\left( \frac{1+e^{-\lambda}}{1-e^{-\lambda}}\right)^{1/2} \hspace{-3mm} -
\frac{e^{-\lambda}\tau^\gamma}{(1+e^{-\lambda})^{1/2}(1-e^{-\lambda})^{3/2}}\,\frac{t^{-\gamma}}{\Gamma(1-\gamma)}
\ee
for $t\to\infty$.  For a slow reaction ($\lambda \to 0$) this reduces to
\be
S^*(t)\sim \sqrt{\frac{2}{\lambda}}-\frac{1}{(2\lambda^3)^{1/2}\Gamma(1-\gamma)}\left(\frac{\tau}{t}\right)^\gamma, \qquad t\to\infty.
\label{longtailslow}
\ee

By comparing Eqs.~(\ref{slreacS}) and (\ref{longtailslow}) we conclude the intuitively obvious result that the decay to the asymptotic value is slower here than in the case of an exponential stepping time distribution since at a given time fewer steps have been taken in the former than in the latter. However, at infinite time the number of steps $S(\infty)$ taken is the same as in the previous case and independent of the details of the waiting time distribution.

\subsubsection{Ultra-slow waiting time distribution}

Finally, we continue to consider reaction events that can only occur concurrently with a jump, but now with an ultra-slow waiting time distribution of the form studied in \cite{DistinctSitesCTRW}:
\be
\psi(t)\sim \frac{\beta \, A}{ t \ln^{\beta+1}{(t/\tau)}}, \qquad t\to\infty.
\label{slowest}
\ee
Here $\beta$ and $A$ are positive.  This long-time behavior translates to the following small $u$ behavior for the Laplace transform of this waiting time distribution,
\be
\label{ultraslowpsi}
\widetilde{\psi}(u)\sim 1-\frac{A}{ \ln^\beta(1/\tau u)}, \qquad u\to 0.
\ee
Substituting this form into Eq.~(\ref{s*u}) we obtain
\be
\widetilde{S}^*(u)\sim
\frac{1}{u}\,
\frac{\left[1+e^{-\lambda}-  e^{-\lambda} A \ln^{-\beta}(1/\tau u)\right]^{1/2}}
{\left[1-e^{-\lambda}+  e^{-\lambda} A \ln^{-\beta}(1/\tau u) \right]^{1/2}}
\ee
for $u\to 0$.

We again implement the power series expansion \eqref{exp1}.
This yields the first two terms in the series for $\widetilde{S}^*(u)$,
\be
\widetilde{S}^*(u)\sim\left(\frac{1+e^{-\lambda}}{1-e^{-\lambda}}\right)^{1/2}\hspace{-3mm}u^{-1}-
\frac{e^{-\lambda} A \ln^{-\beta}(1/\tau u)}{(1+e^{-\lambda})^{1/2}(1-e^{-\lambda})^{3/2}}\,
  u^{-1}
\ee
as $u\to 0$. The usual Tauberian theorem then leads to the temporal behavior at long times,
\be
S^*(t)\sim\left(\frac{1+e^{-\lambda}}{1-e^{-\lambda}}\right)^{1/2}\hspace{-2mm} -
\frac{ e^{-\lambda} A  \ln^{-\beta}(t/\tau)}
{(1+e^{-\lambda})^{1/2}
(1-e^{-\lambda})^{3/2}}
\ee
as $t\to \infty$.
For a slow reaction this leads to the asymptotic result
\be
S^*(t)\sim \sqrt{\frac{2}{\lambda}}-\frac{A}{\sqrt{2\lambda^3}}\, \ln^{-\beta}(t/\tau).
\label{this}
\ee
Comparing this result with Eq.~\eqref{longtailslow} shows that the asymptotic behavior here is approached even more slowly than there.  Thus the approach to the finite value $S^*(\infty)$ is fastest for an exponential waiting time distribution, slower for a long-tailed waiting time distribution, and slowest for an ultra-slow waiting time distribution.

\subsection{Evanescence independent of steps}
\label{notassociated}

In this section we deal with the same three waiting time distributions together with an exponential evanescence for the walkers, but now we consider the case where stepping and evanescence are disconnected, that is, evanescence is now a process independent of the transport of the walkers.  In this case the probability $\rho(t)$ that a walker survives up to time $t$ is straightforwardly given by $\rho(t)=e^{-\lambda t}$.
It should be kept in mind that although we use the same symbol $\lambda$ here and in Eq. \eqref{rhon}, it represents related but different quantities; furthermore, there $\lambda$ has units of (time)$^{-1}\hspace{-1mm}$ whereas it is nondimensional in Eq. \eqref{rhon}.

The mean number of sites visited up to time $t$ is now given by
\be
S^*(t) =\sum_{n=0}^\infty  S_n \chi_n^*(t),
\label{independent}
\ee
where $S_n$ is the average number of sites visited by a walker given that the walker has survived up to the $n^{th}$ step, and $\chi_n^*(t)$ is the probability that the evanescent walker has taken \emph{exactly} $n$ steps up to time $t$. This expression should be compared with Eq.~(\ref{s*t}). Note that here the evanescence is taken into account in the probability that the walker has taken $n$ steps up to time $t$, whereas in the case of evanescence coupled to steps the evanescence is taken into account in the distinct number of sites visited.

In the current situation the distribution of waiting times must take into account the evanescence of the walkers. Explicitly, the probability per unit time that an evanescent walker jumps between $t$ and $t+dt$, $\psi^*(t)$,  is just the waiting time distribution for non-evanescent walkers $\psi(t)$ diminished by a factor $e^{-\lambda t}$, that is, $\psi^*(t) = e^{-\lambda t} \psi(t)$.
In place of Eq.~(\ref{convpsi}) we must now use
\be
\psi^*_{n+1}(t)=\int_0^t \psi^*_n(t')\,\psi^*(t-t')\,dt'.
\ee
Correspondingly, Eq.(\ref{familiar}) must now be replaced by
\be
\label{chi22}
\chi^*_n(t)=\int_0^t \psi^*_n(t')\,\Phi^*(t-t')\,dt',
\ee
where $\Phi^*(t-t')=1-\int_{0}^{t-t'} \psi^*(t'')\,dt''$ is the probability that the walker has not moved
in the time interval $t-t'$ since the $n$-th stepping time $t'$, taking into account that the walker might have been evanesced during this time interval. Laplace transforming this equation gives
\begin{eqnarray}
\widetilde{\chi}^*_n(u)&=&\widetilde{\psi}^*_n(u)\,\Phi^*(u)\nonumber\\
&=&\widetilde{\psi}^*_n(u)\,\frac{1-\widetilde{\psi}^*(u)}{u}\nonumber\\
&=&[\widetilde{\psi}^*(u)]^n\,\frac{1-\widetilde{\psi}^*(u)}{u}\nonumber\\
&=&[\widetilde{\psi}(u+\lambda)]^n\,\frac{1-\widetilde{\psi}(u+\lambda)}{u}.
\end{eqnarray}
Using this in the Laplace transform of Eq.~(\ref{independent}) then yields
\begin{eqnarray}
\label{S*psi}
\widetilde{S}^*(u)&=&\frac{1-\widetilde{\psi}(u+\lambda)}{u}\,\sum_{n=0}^\infty  [\widetilde{\psi}(u+\lambda)]^n\, S_n \nonumber\\
&=&\frac{1-\widetilde{\psi}(u+\lambda)}{u}\,\widehat S(\widetilde{\psi}(u+\lambda)),
\label{this.here}
\end{eqnarray}
where $\widehat S(\cdot)$ is the generating function for the mean number of distinct sites visited by a non-evanescent walker, as given in the first part of
Eq.~(\ref{generating.functions}).

In one dimension we substitute Eq.~(\ref{1dgenfunc}) into Eq.~(\ref{alive}) and this into Eq.~(\ref{this.here}) to arrive at
\be
\label{S*1d}
\widetilde{S}^*(u)=\frac{1}{u}\left(\frac{1+\widetilde{\psi}(u+\lambda)}{1-\widetilde{\psi}(u+\lambda)}\right)^{1/2}.
\ee
Note that, because $\lim_{\lambda\to 0} { \widetilde S}^*(u)=\widetilde S(u)$ and $S^*(0)=S(0)=1$, Eq.~\eqref{S*1d} implies $u \widetilde S^*(u) -S^*(0)= (u+\lambda) \widetilde S(u+\lambda)-S(0)$, or stated differently, we see that  $dS^*/dt=\exp(-\lambda t) dS/dt$.

Let us now define the auxiliary function
\be
\widetilde F(u)
=u \widetilde S(u)
=\left(\frac{1+\widetilde{\psi}(u)}{1-\widetilde{\psi}(u)}\right)^{1/2}.
\ee
Then $S^*(t)=\int_0^t dt' \exp(-\lambda t') F(t')$ or, equivalently,
\be
\label{Sevt1}
S^*(t)=\widetilde F(\lambda)-\int_t^\infty dt' \exp(-\lambda t') F(t'),
\ee
so that
\be
\label{Sinfty1}
S^*(\infty)=\widetilde F(\lambda)=\lambda \widetilde S(\lambda).
\ee

We now proceed to implement our three different waiting time distributions and compare the results obtained here with those obtained for walkers that can only evanesce when they jump.

\subsubsection{Exponential waiting time distribution}
\label{subsec:ewtd}

We again start with the exponential waiting time distribution  (\ref{exponentialwtd}), which immediately gives
\be
\psi^*(t)=\omega e^{-(\omega+\lambda)t}
\ee
when the evanescence process is taken into account. The effective mean waiting time between steps is now $\omega/(\omega+\lambda)^2$, shortened relative to the mean waiting time $T=1/\omega$ for non-evanescent walkers.
In other words, the mean number of steps per unit time is now increased, indicating that those walkers that wait longer to jump are removed with higher probability.
The Laplace transform of $\psi(t)$ is $\widetilde{\psi}(u)=\omega/(\omega+u)$, so that $\widetilde F(u)=[(2\omega +u)/u]^{1/2}$, and from Eq.~\eqref{Sinfty1} one finds
 \be
S^*(\infty) = \left(\frac{2\omega+\lambda}{\lambda}\right)^{1/2},
\ee
to be compared with Eq.~(\ref{asymp1}). Note that the asymptotic value of the distinct number of sites visited there is independent of the stepping rate $\omega$ whereas here that is not the case.  Here there is a competition between the rate of the reaction and the jump frequency.  It is in fact difficult to meaningfully compare the two results because the dependences on the parameters are so different.

The approach to the asymptotic result can be calculated by means of Eq.~\eqref{Sevt1} by approximating $F(t)$ by its long-time behavior. Because
$\widetilde F(u)\sim (2\omega/u)^{1/2}$ for $u\to 0$, one finds $F(t)\sim \sqrt{2\omega/\pi t}$ for long times, and Eq.~\eqref{Sevt1} implies
\begin{equation}
S^*(t)\sim S^*(\infty)-
\sqrt{\frac{2\omega}{\lambda}}\,\text{erfc}\left(\sqrt{-\lambda\,t}\right),\qquad t \to \infty.
\end{equation}
As in Sec.~\ref{associated}, here again the approach to the asymptotic result is faster than exponential.
In the slow reaction limit (small $\lambda$, but still with $\lambda t$ large) the asymptotic result reduces to
\be
S^*(t) \sim \sqrt{\frac{2\omega}{\lambda}} - \frac{1}{\lambda}\,
\sqrt{\frac{2\omega}{\pi t}}\,e^{-\lambda\,t}  , \qquad t \to \infty,
\ee
which is equal to Eq.~(\ref{slreacS}) when here one replaces  $\lambda$ by $\lambda \omega$.

\subsubsection{Long-tailed waiting time distribution}
\label{sub:L-twtd}

We again consider a waiting time distribution which at long times behaves as given in Eq.~(\ref{longtail}),
with Laplace transform for small $u$ as in Eq.~(\ref{asymptpsiu}).
The final value of $S^*(t)$  is then
\be
S^*(\infty)=\lambda \widetilde{S}(\lambda)=\left(\frac{1+\widetilde{\psi}(\lambda)}{1-\widetilde{\psi}(\lambda)}\right)^{1/2}.
\ee
While it is possible to calculate the result for arbitrary $\lambda$, not much is learned from it, so we only exhibit the asymptotic result for a slow reaction.
In this limit we can use the expression for the Laplace transform of the waiting time distribution given in Eq.~(\ref{asymptpsiu}) to calculate $\widetilde{\psi}(\lambda)\sim 1-\tau^\gamma \lambda^\gamma$.  Then
\be
S^*(\infty)\sim \sqrt{\frac{2}{(\tau \lambda)^\gamma}} \quad \mbox{for} \quad \lambda \tau \to 0.
\ee
This behavior is entirely different from that of the corresponding result for walks in which evanescence and jumping are tightly coupled.  The latter asymptotic result is the first term on the right hand side of Eq.~(\ref{longtailslow}). There the asymptotic value depends only on $\lambda$, the rate of evanescence.  Here the result also depends on the parameter $\gamma$ of the waiting time distribution.  The approach to this asymptotic value  for slow evanescence can be found as in Sec.~\ref{subsec:ewtd}: $\widetilde  F(u)\sim 2/\tau^\gamma u^\gamma$ for $u\to 0$, from which it follows that $F(t)\sim 2t^{-1+\gamma}/\tau^\gamma \Gamma(\gamma)$ for large $t$. From Eq.~\eqref{Sevt1} one then gets
\be
S^*(t)  \sim \sqrt{\frac{2}{(\tau\lambda)^\gamma}} - \left(\frac{2}{\tau^\gamma}\right)^{1/2}\frac{1}{\Gamma(\gamma/2)\lambda}\,\frac{e^{-\lambda t}}{t^{1-(\gamma/2)}}
\label{yonder}
\ee
for long times when $\lambda \to 0$ (but still $\lambda t\to \infty$).

It is somewhat difficult to compare this result with that of Eq.~(\ref{longtailslow}). The inverse power of time in the subdominant term  goes as $t^{-\gamma}$ in the coupled case and as $t^{1-(\gamma/2)}$ in the decoupled case.  Which exponent contributes to a more rapid decay depends on the value of $\gamma$, specifically whether it is smaller or larger than $2/3$.  However, at very long times the exponential  term in Eq.\eqref{yonder} is dominant so that eventually near asymptotia the approach is more rapid when evanescence and stepping are decoupled than when they are coupled.

In the absence of evanescence, it has long been known \cite{FirstStepsBook} that the  distinct number of sites visited by an ensemble of walkers (called traps in this context) is related to the survival probability $Q_T(t)$ of an immobile  target located at a given site of an infinite lattice and surrounded by a random distribution of traps of density $c_0$ (= fraction of lattice sites occupied by traps): $Q_T(t)=\exp\{-c_0 [S(t)-1]\}$. This  relation is also valid for evanescent traps, $Q^*_T(t)=\exp\{-c_0 [S^*(t)-1]\}$ \cite{MortalBookChapter}.
This result can also be translated to the continuum limit. The long-time asymptotic behavior of $Q^*_T(t)$ on a one-dimensional lattice for (sub)diffusive traps, which can be modeled as CTRWers with a long-tailed waiting time distribution,  was also studied in  \cite{SurvivalEvanescent1d} by means of a  different approach in which fractional calculus was employed. The results reported in that work [cf. Eq. (19)] fully agree with those obtained here and can also be generalized to higher dimensions \cite{SurvivalEvanescent} via reaction-subdiffusion equations \cite{SokolovSchmidtSaguesPRE06, Zoia2008, AbadYusteLindenberg10, MFHBook, Fedotov13}.

\subsubsection{Ultra-slow waiting time distribution}

We complete our panorama by considering the ultra-slow waiting time distribution  whose long-time behavior is given in Eq.~(\ref{slowest}). Again, for small $\lambda$ it is appropriate to use the small-$u$ form of its Laplace transform as given in Eq.~ \eqref{ultraslowpsi}, to obtain
\begin{eqnarray}
S^*(\infty)&\sim&\left[ 2A^{-1}  \ln^{\beta}{(1/\tau \lambda)}\right]^{1/2} 
\end{eqnarray}
for $\lambda \tau \to 0$. It is again difficult to directly compare this result, valid when the evanescence and jump processes are concurrent, with the first term on the right of Eq.~(\ref{this}), because the parameter dependences are so different.

The approach to the asymptotic limit is again obtained using the same methodology as before, and we find that for small $\lambda$ as $t \to \infty$
\begin{eqnarray}
S^*(t) &\sim& \left[ 2A^{-1}  \ln^{\beta}{(1/\tau \lambda)}\right]^{1/2} \nonumber\\
&&- \beta \, \left( 2A  \right)^{-1/2}
\frac{ \ln^{\beta/2-1}(t/\tau)}{\lambda t} \,e^{-\lambda t}.
\end{eqnarray}
Again, for the comparison of the approach to asymptotic behavior here and in the case of coupled evanescence and jump events, cf. Eq.~(\ref{this}), the remarks following Eq.~(\ref{yonder}) hold here as well, with the remarks about $\gamma$ there translated to the parameter $\beta$ here.

\section{Summary and Conclusions}
\label{conclusions}

Using a CTRW approach we have calculated the asymptotic behavior of the distinct number of sites visited by exponentially evanescing walkers.   We considered two situations, namely one where the evanescence and stepping processes are coupled so that the former does not occur when a walker stands still and waits, but only when it takes a step, and another where these two processes are decoupled.  In the coupled case the density of walkers decreases as $\rho(n) \propto \exp(-\lambda n)$ and we have to be appropriately careful when converting this to a decay in time.  How the density decays with time depends on the waiting time distribution of the walkers.  In the decoupled case the density of walkers decreases as $\rho(t) \propto \exp(-\lambda t)$.  We have presented results for the case of a slow evanescence process, $\lambda \to 0$, because we are mainly interested in the diffusion limit, that is, in the limit where many steps are taken before the walkers on average evanesce (in most cases the discussion makes little sense if the walkers on average evanesce early in the walk).

We next collect our formulas so that they can be assessed when all seen together. We collect the results according to the waiting time distribution.  In all cases we present the number of distinct sites visited as a function of time at long times (and for weak evanescence rate $\lambda$), that is, the asymptotic results and the approach to the asymptotes.  We are of course appropriately cautious in our conversion from step number to time. The first result in each case is that obtained when evanescence and stepping are tightly coupled, the second when they occur independently, and we specifically recall that $\lambda$ does not represent the same quantity in both cases. We also note one additional point before presenting this summary:  When there is no evanescence, it is clear that walkers continue to visit new sites without end, so that the distinct number of sites visited as a function of time diverges as $t\to\infty$ \cite{DistinctSitesCTRW}.  How exactly it diverges depends on the waiting time distribution.  In particular, for an exponential waiting time distribution $\psi(t) = \omega e^{-\omega t}$ we recover the classic one-dimensional random walk result $S(t) \sim \sqrt{8\omega t/\pi}$. When there is evanescence, the distinct number of sites visited approaches a finite limit in all cases.
\begin{itemize}
\item Exponential waiting time distribution
\[\psi(t) = \omega e^{-\omega t}\]
\begin{enumerate}
\item Tightly coupled
\[
S^*(t)\sim\sqrt{\frac{2}{\lambda}}-\frac{1}{\lambda}\sqrt{\frac{2}{\pi \omega t}}\,e^{-\lambda \omega t}
\]
\item Uncoupled
\[
S^*(t) \sim \sqrt{\frac{2\omega}{\lambda}} - \frac{1}{\lambda}\sqrt{\frac{2\omega}{\pi  t}} \, e^{-\lambda t}
\]
\end{enumerate}
\item Long-tailed waiting time distribution
\[\psi(t)\sim \frac{\gamma \tau^\gamma t^{-1-\gamma}}{\Gamma(1-\gamma)}
\]
\begin{enumerate}
\item Tightly coupled
\[
S^*(t)\sim \sqrt{\frac{2}{\lambda}}-\frac{1}{(2\lambda^3)^{1/2}\Gamma(1-\gamma)}\left(\frac{\tau}{t}\right)^\gamma
\]
\item Uncoupled
\[
S^*(t) \sim \sqrt{\frac{2}{(\tau\lambda)^\gamma}} - \left(\frac{2}{\tau^\gamma}\right)^{1/2}\frac{1}{\Gamma(\gamma/2)\lambda}\,\frac{e^{-\lambda t}}{t^{1-(\gamma/2)}}
\]
\end{enumerate}
\item Ultra-slow waiting time distribution
\[\psi(t)\sim \frac{\beta A }{  t [\ln{(t/\tau)}]^{\beta+1}}
\]
\begin{enumerate}
\item Tightly coupled
\[
S^*(t)\sim \sqrt{\frac{2}{\lambda}}-\frac{1}{\sqrt{2\lambda^3}}\,A\,[\ln{(t/\tau)}]^{-\beta}
\]
\item Uncoupled
\begin{eqnarray}
S^*(t) &=& \left( 2 A^{-1} \ln^{\beta}(1/\tau \lambda)\right)^{1/2} \nonumber\\
&&-\left( 2A \right)^{-1/2} \beta \;\frac{ \ln^{\beta/2-1}(t/\tau)}{\lambda t} \,e^{-\lambda t} \nonumber
\end{eqnarray}
\end{enumerate}
\end{itemize}

We point to a result that is intuitively obvious: The asymptotic result $S^*(\infty)$ when the steps and evanescence are tightly coupled is the same regardless of the waiting time distribution.  This is not the case when they are uncoupled.  The approach to the asymptotic result is different for all cases.

As we pointed out at the end of Sec.~\ref{sub:L-twtd},  the  distinct number of sites visited by an ensemble of walkers (called traps  in this context) is related to the survival probability $Q^*_T(t)$ of an immobile  target located at a given site of an infinite lattice surrounded by a random distribution of (evanescent) traps of (initial) density $c_0$ by the equation $Q^*_T(t)=\exp\{-c_0 [S^*(t)-1]\}$.

We also mention an additional interesting connection.  It is well known that so-called anomalous diffusion (which here corresponds to a walk with a non-exponential waiting time distribution in our ``uncoupled" model) may provide an explanation for the observed stretched exponential relaxation (or Kohlrausch-Williams-Watts relaxation) in the so-called defect diffusion model \cite{DefDif1,DefDif2,DefDif3}.  We argued \cite{ourPRL} that the stretched exponential behavior could also occur with ordinary diffusion (which here corresponds to a random walk with an exponential waiting time distribution in our "coupled" model) if the walkers in the model could evanesce. We commented there, and we have confirmed here,  that the combination of anomalous diffusion and evanescence is also a possibility, one that provides a broad array of possible relaxation behaviors.

Routes of future work might include a detailed analysis of the stretched exponential problem, and the extension of our results to higher dimensions, and to different forms of evanescence for both the coupled and uncoupled cases. The full characterization of the exploration properties of one or various evanescent CTRWers via other quantities beyond the mean number of distinct sites visited is also a matter of current interest and should be tractable within our generating function approach.

\begin{acknowledgments}
This work was partially funded by the Ministerio de Ciencia y Tecnolog\'ia (Spain) through Grant No. FIS2010-16587 (partially financed by FEDER funds), by the Junta de Extremadura through Grant No. GRU10158, and by the US National Science Foundation under Grant No. PHY-0855471.
\end{acknowledgments}

\end{document}